\documentclass[12pt]{article}
\usepackage{amssymb}
\usepackage{amsmath}
\usepackage{graphicx}
\def\beq{\begin{equation}}
\def\eeq{\end{equation}}
\usepackage[all,2cell,dvips]{xy}

\def\IR{\relax{\rm I\kern -.18em R}}

\begin{document}
\title{Integrability and BRST invariance from BF topological theory}
\author{ {\Large A. Restuccia$^{1}$ and A. Sotomayor$^{2}$}}
\maketitle{\centerline{$^1$Departamento de F\'{\i}sica,
Universidad de Antofagasta, Chile}}
\maketitle{\centerline {$^2$Departamento de Matem\'{a}ticas,
Universidad de Antofagasta, Chile }}
\maketitle{\centerline{e-mail: alvaro.restuccia@uantof.cl, adrian.sotomayor@uantof.cl}}
\begin{abstract}We consider the BRST invariant effective action of the non-abelian BF topological theory in $1+1$ dimensions with gauge group $Sl(2,\mathbb{R})$. By considering different gauge fixing conditions, the zero-curvature field equation give rise to several well known integrable equations. We prove that each integrable equation together with the associated ghost field evolution equation, obtained from the BF theory, is a BRST invariant system with an infinite sequence of BRST invariant conserved quantities. We construct explicitly the systems and the BRST transformation laws for the KdV sequence (including the KdV, mKdV and CKdV equations) and Harry Dym integrable equation.

\end{abstract}


Pacs: 02.30.lk; 11.15.Ex; 02.30.Jr; 11.30.-j

\section{Introduction}Topological field theories were introduced by Schwarz \cite{Schwarz}, who related the Ray-Singer torsion to the partition function of a quantum field theory, and Witten \cite{Witten}, who introduced the so-called ``Witten index" and opened up a wide range of research lines in mathematics and physics.

Topological BF field theories \cite{Horowitz,Rocek,Myers,Blau} have been extensively studied, in particular in relation to gravity theories in several dimensions \cite{L2,L3,L4,L5,L6,L61,L7,L8,L9,L10,L11,L12,L13}. Two dimensional BF field theories were intensively studied in the context of Poisson sigma models, deformation quantization and nonconmutative field theories.

The BRST formalism (Becchi, Rouet, Stora and Tyutin) \cite{BR1,BR2,BR3,BR4,BR5} is a fundamental approach in quantum field theory, and as such has been used in the analysis of the BF topological theories, both from the perturbative and non perturbative point of view \cite{B1,B2,B3,B4,B5,B6,B7,B8,B9}. The renormalizability of the theory has been proved in \cite{C1,C2,C3,C4,C5,C6,C7}. BF theories have been also used in the description of topological effects in condensed matter
\cite{D1,D2,D3,D4,D5,D6,D7,D8}.

In this work, we study a particular class of solutions of the field equations of the two dimensional BRST formulation of BF theory which are well known integrable equations. This equations arise from the zero curvature field equation in different partial gauge fixing procedures. The origin of these integrable equations as solutions of the zero curvature equation is one side of an old conjecture claimed by Ward \cite{Ward}, that all integrable equations should arise from the selfdual field equation for the curvature of a one-form connection arising from a principal bundle with appropriated structure group, which in particular contains the zero curvature equation. The latter has been extensively studied in the literature. A complete analysis of the zero curvature equation was given in \cite{Fukuyama}.

The main result of our work is that the field equations of the BRST effective action of the two dimensional BF model, in the above mention gauge fixing procedures, extend the integrable equations to BRST invariant integrable systems where together with the well-known integrable equation there is an evolution equation for the ghost field. The new system has an infinite sequence of conserved quantities, which are BRST invariant. The symmetry is explicitly constructed for the KdV sequence of integrable equations as well as for the Harry Dim integrable equation. An interesting point is that the system describes soliton solutions in the ghost sector of the system. The new integrable system, although describes the evolution of fields valued on the even and odd part of a Grassmann algebra as the supersymmetric extension of the KdV equation \cite{H1,H2,H3,H4,H5,H6,H7}, has a different analytical structure and it is directly related to interesting topological problems.

In section 2 we present the notation used in the work. In section 3 we consider de BF action and gauge symmetry of the theory. In section 4 we present details of the BRST effective BF theory, the BRST charge and associated field equations. In section 5 we discuss the evolution equations for the field and its associated ghost field. In sections 6 and 7 we introduce the new integrable BRST systems extending the KdV, mKdV, CKdV and Harry-Dim integrable equations. In section 8 we give our conclusions.
\section{Notation} The generators of the $sl(2,\mathbb{R})$ algebra will be denoted by

\[T_0=\frac{1}{2}\left(\begin{array}{cc} 1 & 0
\\ 0 & -1  \end{array}\right),T_+=\frac{1}{\sqrt{2}}\left(\begin{array}{cc} 0 & 1
\\ 0 & 0  \end{array}\right),T_-=\frac{1}{\sqrt{2}}\left(\begin{array}{cc} 0 & 0
\\ 1 & 0  \end{array}\right). \]
They satisfy \[[T_a,T_b]=f_{ab}^cT_c, a=0,+,-,\]
where the structure contants are antisymmetric on the subindices and satisfy \[f_{0+}^+=1,f_{0-}^-=-1,f_{+-}^0=1,\] with all other components of $f_{ab}^c$, excluding the antisymmetric of the given ones, being zero.

Also,  \[tr(T_aT_b)=\frac{1}{2}\eta_{ab},\eta_{ab}=\eta_{ba},\] 
where  \[\eta_{00}=1,\eta_{0+}=\eta_{0-}=0,\eta_{++}=0,\eta_{+-}=1,
\eta_{--}=0,\]
and $f_{abc}=f_{ab}^d\eta_{dc}=\frac{1}{2}\epsilon_{abc}$, where $\epsilon_{abc}$ is totally antisymmetric with $\epsilon_{0+-}=1.$

The Lie algebra valued connection one-form $A\equiv A_\mu^a T_a dx^\mu,\mu=0,1,$ has curvature two-form $F\equiv dA+A\wedge A= \frac{1}{2}F_{\mu\nu}^aT_adx^\mu\wedge dx^\nu,$ 
hence
\[F_{\mu\nu}^a=\partial_\mu A_\nu^a-\partial_\nu A_\mu^a+f_{bc}^aA_\mu^bA_\nu^c.\]
The local coordinates on ${\mathbb{R}}^2$ will be denoted $x^0=t,x^1=x.$

All the fields are functions of $t$ and $x$, but we only make explicit the $x$ dependance.

\section{The BF theory} We consider the BF topological theory in $1+1$ dimensions valued on the $sl(2,\mathbb{R})$ algebra. 

Its action is given by
\[S(B,A)\equiv \int_{{\mathbb{R}}^2}tr(BF),\] where $B$ is a Lie algebra valued zero-form and $F$ the Lie algebra valued curvature of the connection one-form $A$.

We have
 \begin{eqnarray*}& & tr(BF)=B^aF^btr(T_aT_b)=\frac{1}{2}\eta_{ab}B^aF^b=\frac{1}{2}B_aF^a ,\end{eqnarray*}
 therefore
  \begin{eqnarray*}S(B,A)&=& \frac{1}{4}\int_{{\mathbb{R}}^2}B_aF_{\mu\nu}^a dx^\mu\wedge dx^\nu=
  \\ &=&\frac{1}{2}\int_{{\mathbb{R}}^2}B_a\left(\partial_tA_1^a-\partial_xA_0^a+f_{bc}^aA_0^bA_1^c\right)dt\wedge dx.\end{eqnarray*}

 We may introduce its canonical formulation. The canonical conjugate to $A_1^a$, denoted $\Pi_a$, satisfies the constraint
 \[\Pi_a=\frac{1}{2}B_a.\]
 The canonical Hamiltonian density then can be expressed as
  \begin{eqnarray*}& & \mathcal{H}= -A_0^a\phi_a,\end{eqnarray*}
where $\phi_a
\equiv \partial_x\Pi_a+f_{abc}A_1^b\Pi^c.$

 We have that $\phi_a=0$ is a first class constraint, then
  $\phi_a$ are  the generators of the local $sl(2,\mathbb{R})$ algebra:  \[{\left\{ \phi_a(x),\phi_b(x)  \right\}}_{PB}=f_{ab}^c\phi_c(x)\delta(x-\hat{x}),\] where $f_{ab}^c$ are the structure constants of the $sl(2,\mathbb{R})$ algebra, and we do not write explicitly the $t$ dependance of the fields.
 
  The infinitesimal gauge transformation generated by $\phi_a$ is
 \[\delta_\xi A_1^a(x)=\left\{ \left\langle\xi^c\phi_c\right\rangle,A_1^a(x)  \right\}_{PB}=\partial_x\xi^a(x)+f_{bc}^aA_1^b\xi^c(x)\] where
 $\xi^c\equiv \xi^c(t,x)$ are the infinitesimal parameters of the transformation, and $\langle\rangle$ denotes integration on ${\mathbb{R}}^2$.
 
 The gauge transformation of $\Pi_a$ is
  \[\delta_\xi \Pi_a(x)=\left\{ \left\langle\xi^c\phi_c\right\rangle,\Pi_a(x)  \right\}_{PB}=-f_{ab}^c\Pi_c(x)\xi^b(x)\]
  and of the Lagrange multiplier $A_0^a$,
   \[\delta_\xi A_0^a(x)=\partial_t\xi^a(x)+f_{bd}^aA_0^b(x)\xi^d(x).\]

 \section{The BRST effective action}

 The BRST effective action for BF theories has been already discussed in the literature, as we have refered in the introduction, in particular we follow the approach in \cite{Restuccia1,Restuccia2,Restuccia3}.
 
 The BRST charge associated to the first class constraints $\phi_a=0$ is:
 \[\Omega\equiv \left\langle C^a(x)\phi_a(x)-\frac{1}{2}C^a(x)C^b(x)f_{ab}^d\mu_d(x)   \right\rangle\] where $C^a(t,x)$ are the ghost fields. They are valued on the odd part of a Grassmann algebra and $\mu_a(t,x)$ denote its conjugate momenta.

 They satisfy the following Poisson bracket relations
 \[\left\{ C^a(x),\mu_c(\hat{x})  \right\}_{PB}=\left\{ \mu_c(\hat{x}), C^a(x)  \right\}_{PB}=\delta_c^a\delta(x-\hat{x}).\]

 It follows that
 \beq \left\{ \Omega, \Omega  \right\}_{PB}=0. \label{omega}\eeq
 The BRST transformation of the canonical fieds $A_1^a, \Pi_a,C^a,\mu_a$ is given by
 \[\delta_{BRST}\equiv \zeta\hat{\delta }\] where $\zeta$ is the constant BRST parameter valued on the odd part of a Grassmann algebra.

 The transformation $ \hat{\delta}$ acts as follows
 \begin{eqnarray}\hat{\delta}C^a(x) & \equiv & \left\{ \Omega, C^a(x)  \right\}_{PB}=-\frac{1}{2}C^b(x)C^c(x)f_{bc}^a,
 \\ \hat{\delta}\mu_a(x) & \equiv & \left\{ \Omega, \mu_a(x)  \right\}_{PB}=\phi_a(x)-f_{ab}^dC(x)^b\mu_d(x),
 \\ \hat{\delta}A_1^a(x) & \equiv & \left\{ \Omega, A_1^a(x)  \right\}_{PB}=\partial_xC^a+f_{bc}^aA_1^b(x)C^c(x), \\
 \hat{\delta}\Pi_a(x) & \equiv & \left\{ \Omega, \Pi_a(x)  \right\}_{PB}=-f_{ab}^c\Pi_c(x)C^b(x).      \end{eqnarray}
 It follows from (\ref{omega}) that for the canonical fields  $f_c(x)$  \[\hat{\delta}\hat{\delta}f_c(x)=0.\]

 In fact,
 \[\left\{ \Omega\left\{\Omega,f_c(x) \right\}_{PB}  \right\}_{PB}=\left\{\left\{\Omega,\Omega \right\}_{PB},f_c(x)\right\}_{PB}-
 \left\{ \Omega,\left\{\Omega,f_c(x) \right\}_{PB}  \right\}_{PB}, \] hence $ \left\{ \Omega\left\{\Omega,f_c(x) \right\}_{PB}  \right\}_{PB}=0.$

 We introduce now the BRST transformation for the non-canonical fields, that is, the Lagrange multiplier $A_0^a$ and the antighost field $ \bar{C}^a(x).$ We can also introduce them as canonical fields \cite{BF1,BF2,BF3,BF4,BF5,BF6,BF8,BF9,BF10}, but we consider here a more direct approach \cite{Restuccia3}.

 The transformations are
 \begin{eqnarray*}&& \hat{\delta}\bar{C}_a(x)=D_a(x), \\&& \hat{\delta}D_a(x)=0, \\&& \hat{\delta}A_0^a(x)=\theta^a(x), \\ &&
 \hat{\delta}\theta^a(x)=0,   \end{eqnarray*}
 where $D_a(x)$ and $\theta^a(x)$ are new auxiliary fields.
 
 They satisfy: $\hat{\delta}\hat{\delta}{\bar{C}}_a(x)=\hat{\delta}\hat{\delta}A_0^a(x)=\hat{\delta}\hat{\delta}D_a(x)=\hat{\delta}\hat{\delta}\theta^a(x)=0.$
 
 Hence all the fields in the effective action belong to the kernel of the operator  $\hat{\delta}\hat{\delta}$.

 The effective action for the BF theory is then given by
 \begin{eqnarray}S_{eff}&=&\int_{{\mathbb{R}}^2}\left[\Pi_a\partial_tA_1^a+\mu_a\partial_tC^a+\hat{\delta}\left(A_0^a\mu_a
 \right)+\hat{\delta}\left(\bar{C}_b\chi^b
 \right)\right]dt\wedge dx \nonumber \\ &=& \int_{{\mathbb{R}}^2}\left[\Pi_a\partial_tA_1^a+\mu_a\partial_tC^a+\theta^a\mu_a+A_0^a\hat{\delta}\mu_a+
 D_b\chi^b-\bar{C}_b\hat{\delta}\chi^b\right]dt\wedge dx     \end{eqnarray} where $\chi^a=0$ is the gauge fixing condition.

 We notice that \[\phi_a^{eff}\equiv \hat{\delta}\mu_a=\phi_a-f_{ab}^d\mu_d\] is the effective constraint which extends the classical constraint $\phi_a=0$, in the effective (quantum) BF action. Its associated Lagrange multiplier is $A_0^a$. Besides, $D_b(t,x)$ is the Lagrange multiplier associated to the gauge fixing condition $\chi^a=0.$
 
The effective action $S_{eff}$ is a functional of the fields $A_1^a,
\Pi_a,C^a,\mu_a,A_0^a,\bar{C}_b,\theta^a,D_a$ and it is BRST invariant:
\[\hat{\delta}S_{eff}=0\].
 
 In what follows we will consider a gauge fixing condition
 \[\chi^a=0\] which depends only on the canonical fields $A_1^a(t,x).$
 
 The field equations arising from variations of $S_{eff}$ are the following:
 
 \begin{eqnarray}&& \phi_a^{eff}=0, \\ && F_{01}^a=\partial_tA_1^a-\partial_xA_0^a+f_{bd}^aA_0^bA_1^d=0, \\&& \theta^a=\partial_tC^a+f_{bd}^aA_0^bC^d,\mu_a=0
 \\&& \chi^a=0,\hat{\delta}\chi^a=0,     \end{eqnarray} and
 \beq\partial_t{\Pi}_a=f_{ba}^dA_0^b\Pi_d+D_b\frac{\partial \chi^b}{\partial A_1^a}-\overline{C}_b\frac{\partial(\hat{\delta}\chi^b)}{\partial A_1^a}.\eeq
 Since the action is BRST invariant, its field equations are also BRST invariant. In particular,
 \[\hat{\delta}{F_{01}}^a=f_{bd}^aF_{01}^bC^d,\] hence $F_{01}=0$ implies
 $\hat{\delta}{F_{01}}^a=0.$
 
 \section{Gauge fixing procedure and integrability}
 The zero curvature equations for the $sl(2,\mathbb{R})$ one-form gauge connection $A$ gives rise to several integrable equations as it has been discussed in \cite{Fukuyama}. They follow by taking different partial gauge fixing conditions.
 
 We use the following notation, similar to the one in \cite{Fukuyama} but not equal (the components $A_\mu^a$ are related by $A_\mu^a\rightarrow -A_\mu^a$):
 
 \[A_0^0=2P,A_0^+=\sqrt{2}u,A_0^-=\sqrt{2}Q,A_1^0=2R,A_1^+=\sqrt{2}S,A_1^-=-\sqrt{2}T.\]
 
 The zero curvature equation (8) yields:
 \begin{eqnarray}&& \partial_tR-\partial_xP-uT-QS=0,\\&& \partial_tS-\partial_xu+2PS-2uR=0, \\&& \partial_tT+\partial_xQ-2PT-2QR=0.   \end{eqnarray}
 We impose the partial gauge fixing condition \cite{Fukuyama}
 \[R=0,S=1.\]
 
 It follows from (12),(13) and (14) that 
 \begin{eqnarray}&& Q=-\partial_x P-uT, \\&& P=\frac{1}{2}\partial_xu,\\&&
  \Upsilon \equiv \partial_tT-\frac{1}{2}\partial_{xxx}u-
  \partial_x(uT)-(\partial_x u)T=0.\end{eqnarray}
  The zero-curvature condition reduces then to eq. (17).
  
  Besides, the BF field equations (10) imply
  \begin{eqnarray}&& \chi^0=A_1^0=0,\chi^+=A_1^+-\sqrt{2}=0,\\&&C^0=\frac{1}{\sqrt{2}}\partial_xC^+,\\&&C^-=-TC^+-\frac{1}{2}\partial_x\partial_x C^+,   \end{eqnarray}
  that is, the ghost fields reduce solely to $C^+(t,x)$.
  
  (17) is invariant under the BRST transformations, in fact,
  \[\hat{\delta}\Upsilon=2\Upsilon\partial_x\widetilde{C}^++(\partial_x\Upsilon)\widetilde{C}^+,\widetilde{C}^+\equiv\frac{C^+}{\sqrt{2}},\]
  while the BRST transformations of $u$ and $T$ are
  \begin{eqnarray}&&\hat{\delta}u=\partial_t\widetilde{C}^+-u\partial_x \widetilde{C}^++(\partial_xu)\widetilde{C}^+\\&&\hat{\delta}T=
  \frac{1}{2}\partial_{xxx}\widetilde{C}^++(\partial_xT)\widetilde{C}^++
  2T\partial_x\widetilde{C}^+.\end{eqnarray}
  We notice that (21) and (22) are the infinitesimal transformations obtained in \cite{Fukuyama} for $u$ and $T$ in terms of the field $y(t,x)$ when we replace $\widetilde{C}^+\rightarrow \epsilon y$. 
  
  In distinction to the infinitesimal transformations in \cite{Fukuyama}, (21) and (22) are exact transformations under which (17) is invariant.
  
  In \cite{Fukuyama} they consider also a further gauge restriction $T=\frac{u^\alpha}{s},$ where $\alpha$ and $s$ are real numbers. (17) reduces then to
  \beq \partial_t u=\frac{\alpha+2}{\alpha}uu_x+\frac{s}{2\alpha}u^{1-\alpha}u_{xxx},   \eeq
  and \beq \hat{\delta}T-\hat{\delta}\left(\frac{u^\alpha}{s}\right)=0   \eeq implies, using (21) and (22), the following equation for the ghost
  $\widetilde{C}^+$:
  \beq \partial_t\widetilde{C}^+=\frac{s}{2\alpha}
  u^{1-\alpha}\partial_{xxx}\widetilde{C}^+
  +\left(\frac{2}{\alpha}+1\right)u\partial_x\widetilde{C}^+.  \eeq
  Equations (23) and (25) become invariant under the following BRST transformation
  \begin{eqnarray}&&\hat{\delta}\widetilde{C}^+=\widetilde{C}^+\partial_x\widetilde{C}^+,\\&&\hat{\delta}u=\frac{s}{2\alpha}u^{1-\alpha}\partial_{xxx}\widetilde{C}^++\widetilde{C}^+\partial_xu+\frac{2}{\alpha}u\partial_x\widetilde{C}^+,   \end{eqnarray}
  where (26) arises from (2) and (19),(20) while (27) follows directly from (21),(22) and (24). This is an exact symmetry of the evolution eqs. (23) and (25).
  
  We notice that the transformation law (26) is only present in a BRST formulation.
  
  Eq. (23) for $\alpha=1,s=2$ becomes the KdV equation and for $\alpha=-2,s=1$ is the Harry Dym integrable equation. In both cases, given a solution $u(t,x)$ of (23) one can construct an algorithmic procedure to obtain an infinite sequence of solutions of the ghost equations and from them the known infinite sequence of conserved quantities for both integrable systems \cite{Fukuyama}.
  
  In \cite{BF10} it was shown that for any conserved quantity $H$ of an evolution equation $u_t=K(u,u_x,\ldots)$, where dotes denote higher order derivative terms with respect to $x$ up to $m$ derivatives,
  \[H=\int_{-\infty}^{+\infty}\mathcal{H}(u,u_x,\ldots)dx\]
  the gradient $\mathcal{M} \mathcal{H},$
  \[\mathcal{M}\mathcal{H}\equiv\sum_{r=0}^{m}{(-1)}^i\partial_x^i
\frac{\partial\mathcal{H}}{\partial a_i},a_i\equiv \partial_x^iu,\]
satisfies the integral equation

\beq \int_{-\infty}^{+\infty}\left(\frac{\partial (\mathcal{M}\mathcal{H})}{\partial t}\delta u+dK(u,\delta u) \mathcal{M}\mathcal{H}\right)dx=0.\eeq
It turns out that for $\alpha=1,s=2$ (KdV) eq. (28) becomes exactly the evolution equation for the ghost field. This is not the case for other values of $\alpha$. In order to generalize this result for any $\alpha$, it is better to redefine the gauge fixing condition as
\[u=sT^\beta,\]
and rewrite the system (23),(25) in terms of $T$ and $\widetilde{C}^+$.

We end up with the following system
\begin{eqnarray}&&\partial_tT=\frac{s}{2}\partial_{xxx}\left(T^\beta\right)+\left(2\beta+1\right)sT^\beta\partial_xT\\&&\partial_t\widetilde{C}^+=\frac{s}{2}\beta T^{\beta-1}\partial_{xxx}\widetilde{C}^++s\left(2\beta+1\right)T^\beta\partial_x\widetilde{C}^+.    \end{eqnarray}
If $ H=\int_{-\infty}^{+\infty}\mathcal{H}(u,u_x,\ldots)dx$ is a conserved quantity under the evolution equation (29) then the gradient $\mathcal{M}\mathcal{H}$ satisfies the ghost eq. (30) for any value of $\beta,s$. In particular, for $\beta=1,s=2$ (KdV) and for $\beta=\frac{1}{2},s=1$ (Harry Dym).

\section{The integrable BRST system} In this section we show that the system (29),(30) for $\beta=1,s=2$ and for $\beta=\frac{1}{2},s=1$ are integrable systems with an infinite sequence of conserved quantities which are BRST invariant. That is, the KdV and the Harry Dym equations together with their associated ghost evolution equation have an infinite sequence of BRST invariant conserved quantities.

The system (29),(30) is invariant under the BRST transformations (22) and (26). We notice that the conserved quantities of KdV or Harry Dim equations are not BRST invariant.

For example, \[H_0=\int_{-\infty}^{+\infty}Tdx\] is a conserved quantity of the evolution equation (29) for any $\beta$ and $s$. 

However, it is not BRST invariant under (22):
\[\hat{\delta}H_0=\int_{-\infty}^{+\infty}\hat{\delta}T=
\int_{-\infty}^{+\infty}T\partial_x\widetilde{C}^+\neq0.\]
Let \beq H_n=\int_{-\infty}^{+\infty}\mathcal{H}_n\left(T,T_x,\ldots\right)dx\eeq
be a conserved quantity of (29). 

It then follows the existence of $\mathcal{J}_n$ such that
\[\partial_t\mathcal{H}_n=\partial_x\mathcal{J}_n.\]
Consequently,
\[\partial_t\hat{\delta}\mathcal{H}_n=\hat{\delta}\partial_t\mathcal{H}_n=\hat{\delta}\partial_x\mathcal{J}_n=\partial_x\hat{\delta}\mathcal{J}_n\]
which implies that \beq \int_{-\infty}^{+\infty}\hat{\delta}\mathcal{H}_ndx    \eeq is a conserved quantity of the system (29),(30). Moreover, it is invariant under (22),(26).

In particular, $H_0=\int_{-\infty}^{+\infty}T$ has an associated BRST invariant conserved quantity
\beq\widetilde{H_0}=\int_{-\infty}^{+\infty}\hat{\delta}Tdx=\int_{-\infty}^{+\infty}T\partial_x\widetilde{C}^+.\eeq
For each conserved quantity of (29) there is a BRST invariant conserved quantity of (29),(30).

It then follows that for each integrable equation arising from a partial gauge fixing of the one-form connection with zero curvature there is an integrable system, BRST invariant, describing also the evolution of the associated ghost field. The system has an infinite sequence of BRST invariant conserved quantities given by (32).

Another conserved quantity of (29) for any $\beta$ and $s$ is
\beq H_1=\int_{-\infty}^{+\infty}T^{\beta+1}dx.   \eeq

The associated BRST conserved quantity is
\beq\widetilde{H_1}= \int_{-\infty}^{+\infty}\left(\beta+1\right)T^\beta\left(\frac{1}{2}\partial_{xxx}\widetilde{C}^++(\partial_xT)\widetilde{C}^++2T\partial_x\widetilde{C}^+\right).      \eeq
In \cite{Fukuyama} several integrable equations were obtained: KdV, modified KdV, sine-Gordon, Harry Dim, Calogero KdV, nonlinear Schr\"{o}dinger and KdV sequences. To each of them one can construct, following the BRST approach we have considered, a ghost field evolution equation. The system is BRST invariant and has an infinite sequence of conserved quantities.

For the BRST invariant KdV system (29),(30) with $\alpha=1,s=2,$ the first few conserved quantities, where we use $\partial_xg\equiv g_x$, are 
\[H_1=\int_{-\infty}^{+\infty}\hat{\delta}udx,H_3=\int_{-\infty}^{+\infty}\hat{\delta}\left(\frac{1}{2}u^2\right)dx,H_5=\int_{-\infty}^{+\infty}\hat{\delta}\left[\frac{1}{3}u^2-\frac{1}{3}{(u_x)}^2\right]dx\]
which have the explicit expressions
\begin{eqnarray*}&&H_1=\int_{-\infty}^{+\infty}u\widetilde{C}_x^+dx,\\&&
	H_3=\int_{-\infty}^{+\infty}\left(u{\widetilde{C}^+}_{xxx}
	+\frac{3}{2}u^2\widetilde{C}_x^+ \right)dx,\\&& H_5=\int_{-\infty}^{+\infty}\left(\frac{2}{3}u_{xx}{\widetilde{C}^+}_{xxx}+\frac{5}{3}u^3{\widetilde{C}^+}_x
	+u^2{\widetilde{C}^+}_{xxx}+\frac{1}{3}{(u_x)}^2{\widetilde{C}^+}_x-\frac{4}{3}uu_{xxx}\widetilde{C}^+\right)dx.\end{eqnarray*}
Since KdV equation has an infinite sequence of conserved quantities, which are not BRST invariant, one can construct an infinite sequence of independent BRST invariant conserved quantities of the KdV and ghost field system.

\section{The BRST invariant KdV sequence}
We now extend the KdV sequence of integrable equations: KdV,mKdV, CKdV,... to BRST invariant systems with an infinite sequence of BRST conserved quantities.

We impose the partial gauge fixing conditions
\begin{eqnarray}&&A_1^+=\sqrt{2},\\&&A_1^-=0.      \end{eqnarray}

Using the same notation for the components of the connection one form $A_\mu^a$ as in section 5, the zero-curvature field equations became
\beq 2\left(\partial_x-2R\right)\left(R_t-P_x\right)\equiv 2{\left(R_x-R^2\right)}_t-u_{xxx}-4\left(R_x-R^2\right)u_x-2{\left(R_x-R^2\right)}_xu=0,   \eeq
where $P=\frac{1}{2}u_x+uR.$

If we impose, further on, the gauge fixing condition (Miura transformation)
\beq u=2\left(R_x-R^2\right),    \eeq
eq. (38) yields the KdV equation for $u$. 

Moreover, as it is well known,
\beq R_t-P_x\equiv R_t-R_{xxx}+6R^2R_x=0     \eeq is the mKdV equation. If $R$ is a solution of mKdV then $u$ given in (39) is a solution of the KdV equation.

From $\hat{\delta}A_1^+=0,\hat{\delta}A_1^-=0,$ using the BRST transformation law in section 3, we obtain
\begin{eqnarray}&&C^0=\left(\partial_x+2R\right)\widetilde{C}^+\\&&\left(\partial_x-2R\right)\widetilde{C}^-=0.\end{eqnarray} 
Finally,
\beq \hat{\delta}u=2\hat{\delta}\left(R_x-R^2\right)      \eeq       
yields the evolution equation for the ghost field $\widetilde{C}^+,
$
\beq {\widetilde{C}^+}_t={\widetilde{C}^+}_{xxx}+6\left(R_x-R^2\right)\widetilde{C}^+.     \eeq
The BRST transformation for $R$ and $\widetilde{C}^+$ are
\begin{eqnarray}&&\hat{\delta}R=\frac{1}{2}{\widetilde{C}^+}_{xx}+{\left(R\widetilde{C}^+\right)}_x+\widetilde{C}^-,\widetilde{C}^-=\frac{C^-}{\sqrt{2}}\\&&\hat{\delta}\widetilde{C}^+=\widetilde{C}^+{\widetilde{C}^+}_x     \end{eqnarray}
respectively.

(45) and (46) define a BRST transformation under which the mKdV and associated ghost field given by (44) remain exactly invariant. The system mKdV (44) has an infinite sequence of BRST conserved quantities obtained by the action of the BRST operator $\hat{\delta}$ to the conserved quantities of the mKdV equation.

Equation (44), with the identification $u=2(R_x-R^2)$, is the same ghost evolution equation obtained for the KdV equation.

Also (44) is the same BRST transformation law for the ghost field of the KdV eq.

The next integrable equation in the KdV sequence is the CKdV equation. We consider the partial gauge fixing (36),(37) together with
\beq R=v+w,    \eeq where $v$ satisfies the mKdV equation, and the gauge fixing condition
\beq u=2\left(v_x-v^2\right),    \eeq where $v$ satisfies the mKdV equation.

We get
\beq R_x-R^2=v_x-v^2+\left(w_x-2vw-w^2\right),   \eeq
hence, if \beq w_x-2vw-w^2=0,    \eeq then, from (38), $u=2(v_x-v^2),$ satisfies the KdV equation. Moreover (40) yields
\beq w_t-w_{xxx}+\frac{1}{2}\left(w^3+3{w_x}^2w^{-1}\right)=0,  \eeq 
the CKdV equation.

If $w$ satisfies the CKdV equation then $v$ obtained from (50) satisfies the mKdV equation.

The ghost field equations are now given by
\begin{eqnarray*}&&C^0={\widetilde{C}^+}_x+2\left(v+w\right)\widetilde{C}^+\\&&\left[\partial_x-2\left(v+w\right)\right]C^-=0\end{eqnarray*}
and the evolution equation for the ghost field $\widetilde{C}^+$ is
\beq {\widetilde{C}^+}_t={\widetilde{C}^+}_{xxx}+6\left(v_x-v^2\right){\widetilde{C}^+}_x.\eeq
The BRST transformation law for $w$ and $\widetilde{C}^+$ are
\[\hat{\delta}w={\left(w\widetilde{C}^+\right)}_x+{\widetilde{C}^-}_w,\]
where ${\widetilde{C}^-}_w=\widetilde{C}^--{\widetilde{C}^-}_v,$
\[\left[\partial_x-2v\right]{\widetilde{C}}^-_v=0, \]
while the BRST transformation of $\widetilde{C}^+$ is given by (46).

We notice that (46) and (52) are the same as the ones for the BRST mKdV system and for the BRST KdV system. The other members of the KdV sequence follow from the same gauge fixing conditions (36),(37),(48) by considering \[R=v+w+z+\ldots,\]
using the generalization of (49),(50) and the evolution equation (40).

All the members of the KdV sequence share the same ghost evolution equation for $\widetilde{C}^+$ and the same BRST transformation for $\widetilde{C}^+$.

\section{Conclusions}We considered the non-abelian topological model in $1+1$ dimensions with gauge group $Sl(2,\mathbb{R})$. The corresponding BRST formulation of the effective action yields a BRST invariant system of field equations. In particular, the zero-curvature equation and the associated ghost field equations. Using different gauge fixing procedure introduced in \cite{Fukuyama} we obtained BRST invariant systems of evolution equations. They are the BRST extension of the integrable equations arising from the zero-curvature condition. We explicitly performed the construction for the KdV BRST sequence of integrable equations. We proved that each BRST invariant system of the sequence has an infinite sequence of BRST invariant conserved quantities.

We also obtained the BRST invariant system associated to the Harry Dym integrable equation.

From the zero-curvature condition one can obtain several integrable equations. In particular, the KdV sequence, Harry Dym, Sine Gordon, non-linear Schr\"{o}dinger. 

Each integrable equation has an associated evolution equation for the ghost field. The system is BRST invariant and it has an infinite sequence of conserved quantities which are BRST invariant. All of them are then integrable systems preserved under the BRST transformation.

An interesting point to be discussed elsewhere is the geometrical interpretation of the soliton solutions of the field equations in the context of the BF topological field theory.

\section{Conflict of interest} The authors state that there is no conflict of interest.

\end{document}